\RequirePackage{lineno}
\documentclass[5p]{elsarticle} 

\usepackage{atlasphysics}
\usepackage{subfigure}
\usepackage{xspace}
\usepackage{multirow}
\usepackage[hyperindex,breaklinks]{hyperref} 
\usepackage{hypernat}
\usepackage{graphicx}
\usepackage{tabularx}

\newcommand{\mmess}{\ensuremath{M_{\mathrm{mess}}}\xspace}
\newcommand{\mtrans}{\ensuremath{m_{\mathrm{T}}}\xspace}
\newcommand{\Wjets}{\ensuremath{\Wboson+\mathrm{jets}}\xspace}
\newcommand{\Zjets}{\ensuremath{\Zboson+\mathrm{jets}}\xspace}

\renewcommand*{\today}{June 19, 2012}

\usepackage{preprintcover}  

\PreprintCoverPaperTitle{Search for supersymmetry with jets, missing transverse momentum and at least one hadronically decaying $\tau$ lepton in proton-proton collisions at $\sqrt{s}=7\tev$ with the ATLAS detector}  

\PreprintIdNumber{CERN-PH-EP-2012-076}  

\PreprintCoverAbstract{A search for production of supersymmetric particles in final
states containing jets, missing transverse momentum, and at least one
hadronically decaying $\tau$ lepton is presented. The data were recorded by the ATLAS
experiment in $\sqrt{s}=7\tev$ proton-proton collisions at the Large
Hadron Collider. No excess above the Standard Model background
expectation was observed in $2.05\ \ifb$ of data. The results are interpreted
in the context of gauge mediated supersymmetry breaking models with $\mmess=250\tev$, $N_5=3$, $\mu>0$, and
$C_\mathrm{grav}=1$. The production of supersymmetric particles
is excluded at 95\% C.L. up to a supersymmetry breaking scale $\Lambda=30\tev$, independent
of $\tan\beta$, and up to $\Lambda=43\tev$ for large $\tan\beta$.}  

\PreprintJournalName{Physics Letters B}

\begin{document}

\begin{frontmatter}

\title{Search for supersymmetry with jets, missing transverse momentum and at least one hadronically decaying $\tau$ lepton in proton-proton collisions at $\sqrt{s}=7\tev$ with the ATLAS detector}

\author{The ATLAS Collaboration}

\date{\today}

\begin{abstract}
  A search for production of supersymmetric particles in final
states containing jets, missing transverse momentum, and at least one 
hadronically decaying $\tau$ lepton is presented. The data were recorded by the ATLAS
experiment in $\sqrt{s}=7\tev$ proton-proton collisions at the Large
Hadron Collider. No excess above the Standard Model background
expectation was observed in $2.05\ \ifb$ of data. The results are interpreted
in the context of gauge mediated supersymmetry breaking models with $\mmess=250\tev$, $N_5=3$, $\mu>0$, and
$C_\mathrm{grav}=1$. The production of supersymmetric particles
is excluded at 95\% C.L. up to a supersymmetry breaking scale $\Lambda=30\tev$, independent
of $\tan\beta$, and up to $\Lambda=43\tev$ for large $\tan\beta$.

\end{abstract}


\journal{Physics Letters B}  

\end{frontmatter}

\section{Introduction}

Supersymmetry (SUSY)~\cite{Miyazawa:1966,Ramond:1971gb,Golfand:1971iw,Neveu:1971rx,Neveu:1971iv,Gervais:1971ji,Volkov:1973ix,Wess:1973kz,Wess:1974tw} is a well-motivated theoretical concept that introduces a symmetry between bosons and fermions. As a consequence, every Standard Model (SM) particle has a SUSY partner with the same mass and quantum numbers except for the spin which differs by half a unit.  Since none of these partners has been observed SUSY must be a broken symmetry if realised in nature.  If R-parity is conserved~\cite{Fayet:1976et,Fayet:1977yc,Farrar:1978xj,Fayet:1979sa,Dimopoulos:1981zb}, SUSY particles can only be produced in pairs and would decay through cascades involving lighter SUSY particles.  These decay cascades end in the production of the lightest supersymmetric particle (LSP), which is stable and escapes the detector unseen, giving rise to missing transverse momentum in the detector. SUSY can remedy various shortcomings of the Standard Model, such as the hierarchy problem~\cite{Dimopoulos:1981zb,Witten:1981nf,Dine:1981za,Dimopoulos:1981au,Sakai:1981gr,Kaul:1981hi}, the lack of a dark matter candidate~\cite{Goldberg:1983nd,Ellis:1983ew} and the non-unification of the gauge couplings~\cite{Giunti:1991ta,Ellis:1990wk,Amaldi:1991cn,Langacker:1991an}. To achieve this, the masses of at least some SUSY particles must be near the weak scale, and therefore, if weak-scale SUSY is realised in nature, there are good prospects to discover it at the Large Hadron Collider (LHC).

In certain SUSY models, large mixing between left and right sfermions, the partners of the
left-handed and right-handed SM fermions, implies that the lightest sfermions 
belong to the third generation. This leads to a large production rate of $\tau$ leptons 
from decays of $\tilde{\tau}$ sleptons and gauginos, the partners of the SM gauge bosons, 
in SUSY cascade decays. 
For example, 
in the context of Gauge Mediated SUSY Breaking (GMSB)~\cite{Dine:1981gu,AlvarezGaume:1981wy,Nappi:1982,Dine:1993yw,Dine:1994vc,Dine:1995ag} the lighter of the two $\tilde{\tau}$ sleptons 
is the next-to-lightest supersymmetric particle (NLSP) for a large part of the parameter space,
and the very light gravitino, $\tilde{G}$, is the LSP. 
Hence $\tilde{\tau}$ sleptons decay to a $\tau$ lepton and a gravitino.
While this $\tilde{\tau}\to\tau \tilde{G}$ process is the dominant source of $\tau$ leptons from SUSY decays
in certain regions of GMSB model parameter space, the analysis presented here is sensitive
to any process producing $\tau$ leptons in association with jets and
missing transverse momentum.

This Letter presents a search for supersymmetry in final states with at least one hadronically decaying $\tau$ lepton,
missing transverse momentum and jets with the ATLAS detector at the LHC. 
The results of the search 
are interpreted within the GMSB model.
Previous experiments at LEP~\cite{opal, aleph, delphi} have placed constraints on $\tilde{\tau}$ and
$\tilde{e}$ masses and on more generic GMSB signatures. Among these the limits from the OPAL experiment~\cite{opal}
were the most stringent, excluding $\tilde{\tau}$ NLSPs with masses below $87.4 \gev$.
The D0 Collaboration performed a search for squark production
in events with hadronically decaying $\tau$ leptons, jets, and missing
transverse momentum~\cite{dzero_tau}, and the
CMS Collaboration performed searches for new physics in same-sign ditau events~\cite{cms_ditau} 
and multi-lepton events~\cite{cms_multilep} including $\tau$ pairs, but the GMSB model was not specifically considered in any of these results.
A search for supersymmetry in final states containing at least two hadronically
decaying $\tau$ leptons, missing transverse momentum, and jets with the
ATLAS detector is presented in another Letter~\cite{atlas_ditau}.

\section{ATLAS detector}

The ATLAS detector~\cite{ATLASDetector}
is a multipurpose particle physics apparatus with a forward-backward
symmetric cylindrical geometry and nearly $4\pi$ coverage in solid
angle~\footnote{ATLAS uses a right-handed coordinate system with
its origin at the nominal interaction point in the centre of the
detector and the $z$-axis coinciding with the axis of the beam
pipe. The $x$-axis points from the interaction point to the
centre of the LHC ring, and the $y$-axis points upward.
Cylindrical coordinates $(r,\phi$) are used in the transverse
plane, $\phi$ being the azimuthal angle around the beam pipe. The
pseudorapidity is defined in terms of the polar angle $\theta$
as $\eta=-\ln\tan(\theta/2)$}.
The inner tracking detector consists of a silicon pixel detector,
a silicon microstrip detector, and a transition radiation tracker.
The inner detector is surrounded by a thin superconducting solenoid
providing a $2 \, \rm T$ axial magnetic field and by high-granularity liquid-argon
sampling calorimeters. An iron-scintillator tile calorimeter provides
hadronic coverage in the central rapidity range. A muon spectrometer
consisting of large superconducting toroids and 
a system of precision tracking chambers
surrounds the calorimeters.

\section{Data and simulated samples}

The analysis is based on data collected by the ATLAS detector in
proton-proton collisions at a centre-of-mass energy of $7\tev$ between
March and August 2011. Application of beam, detector, and data-quality
requirements resulted in an integrated luminosity of $2.05 \pm 0.08~\ifb$~\cite{lumi, lumi2}.
The data were collected using triggers based
on one jet with transverse momentum $\pt > 75 \gev$, measured at the raw electromagnetic scale,
and missing transverse momentum above $45 \gev$.

In GMSB models, the breaking of SUSY is mediated through flavour-blind SM gauge interactions
of messenger fields with mass scale \mmess which is small compared to the Planck mass.
In addition to \mmess, the free parameters in GMSB models are 
the scale of the SUSY breaking, $\Lambda$,
the number of messenger fields, $N_5$, 
the sign of the Higgsino mixing parameter, $\mathrm {sign}(\mu)$,
the scale factor for the gravitino mass, $C_\mathrm{grav}$, 
and the ratio of the vacuum expectation values of the two Higgs doublets, $\tan\beta$.
In this analysis, GMSB models are studied in the $\Lambda-\tan\beta$ plane for fixed
$\mmess=250\tev$, $N_5=3$, $\mathrm {sign}(\mu)=+1$ and $C_\mathrm{grav}=1$. 
The chosen set of parameter values
restricts the analysis to specific final states relevant for the search with $\tau$ leptons
and to promptly decaying NLSPs. For $N_5 \ge 2$ and large $\tan\beta$ the lightest
$\tilde{\tau}$ slepton, $\tilde{\tau}_1$, is the NLSP. 

Samples of simulated GMSB events are generated with the \texttt{Herwig++}~\cite{Herwigplusplus}
generator for ten values of $\Lambda$ in the range $10 \le \Lambda \le 85\tev$ and 
ten values of $\tan\beta$ in the range $2 \le \tan\beta \le 45$, with the SUSY mass spectra generated using \texttt{ISAJET 7.80}~\cite{isajet}.
The \texttt{MRST2007~LO*}~\cite{MRST2007LO} parton distribution functions (PDFs) are used. 
The production cross sections are calculated with \texttt{PROSPINO}~\cite{Prospino1, Prospino2, Prospino3, Prospino4} 
to next-to-leading order in the QCD coupling using the next-to-leading-order \texttt{CTEQ6.6}~\cite{CTEQ66} PDF set.
The two samples with $\Lambda = 30 \ (40) \tev$ and $\tan\beta=20 \ (30)$, which have cross sections
of $1.95 \ (0.41)$~pb, are used as representative 
points for the optimization of the event selection.

The dominant background processes in this search are  production of 
$\Wboson$ and $\Zboson$ bosons in association with jets (\Wjets and \Zjets),
top quark pair ($t\bar{t}$) and single top quark production.
The \Wjets and \Zjets production processes are simulated 
with the \texttt{ALPGEN}~\cite{Alpgen} generator, using the \texttt{CTEQ6L1}~\cite{CTEQ6L1} PDF set, and are 
normalized to a cross section of $31.4\ \mathrm{nb}$
 and $9.02\ \mathrm{nb}$~\cite{Hamberg:1990np,Melnikov:2006di,Melnikov:2006kv}, respectively. 
The \ttbar, single-top and diboson production processes are generated 
with \texttt{MC@NLO}~\cite{MCatNLO} and the \texttt{CTEQ6.6}~\cite{CTEQ66} PDF set, and are normalized 
using a cross section of $0.165\ \mathrm{nb}$, $0.085\ \mathrm{nb}$~\cite{Bonciani:1998vc,Beneke:2009ye,Moch:2008qy}
and $0.071\ \mathrm{nb}$~\cite{Campbell:1999ah,Campbell:2011bn}, respectively.  
Parton showers and hadronization are simulated with 
\texttt{HERWIG} and the underlying event is modelled with \texttt{JIMMY}~\cite{JIMMY}.
The programs \texttt{TAUOLA}~\cite{TAUOLA,TAUOLA2} and \texttt{PHOTOS}~\cite{PHOTOS} are used to 
model the decays of $\tau$ leptons and the radiation of photons,
respectively.
The production of multijet events is simulated with \texttt{PYTHIA}~\cite{Pythia},
though the multijet background yield in this analysis is estimated using data.
All simulated samples are processed through a full simulation 
of the ATLAS detector~\cite{ATLASSIM} based on \texttt{G\sc{eant4}}~\cite{Geant4}. 
To match the pile-up (overlap of several interactions 
in the same bunch crossing) observed in the data,
the generated signal and background events are overlaid with
minimum-bias events~\cite{pileup1,pileup2} 
and the resulting events are reweighted so that
the distribution of the number of interactions
per bunch crossing agrees with the data.

\section{Object reconstruction}

Jet candidates are reconstructed with the anti-$k_t$ clustering
algorithm~\cite{antikt} with radius parameter $R=0.4$. The inputs
to this algorithm are clusters of calorimeter cells seeded by
cells with energy significantly above the measured noise. Jets
are constructed by performing a four-vector sum over these clusters,
treating each cluster as a four-vector with zero mass. Jets are
corrected for calorimeter non-compensation, upstream material,
and other effects using \pt- and \eta-dependent correction factors
obtained from Monte Carlo simulation and validated with extensive test-beam
and collision-data studies~\cite{jetcalib}. Only jet candidates with
$\pt>30\gev$, $\abseta<2.8$ and a distance $\Delta R>0.2$ with respect to the 
nearest identified electron are considered as real hadronic jets,
where the distance is defined as $\Delta R=\sqrt{(\Delta\eta)^2+(\Delta\phi)^2}$.

The electron and muon identification criteria are identical to those
in Ref.~\cite{onelep}. Electrons and muons are only considered if they
satisfy $\pt>20\gev$ and $\Delta R>0.4$ with respect to the nearest identified jet.

The magnitude of the missing transverse momentum, \met, is computed from the vector
sum of the transverse momenta of all identified electrons and muons,
all jets, and remaining clusters of calorimeter cells with $\abseta<4.5$~\cite{MET}.

Hadronically decaying $\tau$ leptons are
reconstructed from jet candidates with $\pt>10\gev$
and are distinguished from quark- or gluon-initiated jets using a boosted
decision tree (BDT) based on eleven discriminating shower-shape and tracking
variables~\cite{tauid}. Electrons are further rejected using
transition radiation and calorimetric information.
An energy calibration factor for hadronically decaying $\tau$
leptons is applied as function of \pt\ and \eta.
Candidates are required
to satisfy $\pt^\tau>20\gev$ and $\abseta<2.5$ and to have one or three
associated reconstructed tracks (prongs) with total charge $\pm 1$.
The $\tau$ candidates are required to satisfy a \pt-dependent
BDT output criterion~\cite{tauid}
chosen to give $\sim 30\%$ ($\sim 50\%$) signal efficiency
for one-prong (three-prong) $\tau$ candidates as estimated in
$\Zboson(\to\tau\tau)+\mathrm{jets}$ events.
The BDT selection has a corresponding
background acceptance of $\sim 0.5\%$ ($\sim 3\%$), estimated in dijet events,
and the different selection criteria reflect different abundances
of one- and three-prong jets in background samples.

During a part of the data-taking period, an electronics failure in the
liquid-argon calorimeter created a dead region in the second and third layer
of the calorimeter,
corresponding to approximately $1.4\times 0.2\ \mathrm{rad}$ in
$\Delta\eta\times\Delta\phi$. 
A correction is made to the jet energy
using energy depositions in cells neighbouring the dead region; events
having at least one jet, including the leading $\tau$ candidate, 
in this region for which the corrected energy is above $30\gev$ are discarded, 
resulting in a loss of $\sim 6\%$ of the data sample.

\section{Event selection}\label{sec:evsel}

Events are required to have a reconstructed primary vertex with at least
five associated tracks with $\pt > 500\MeV$. Events are rejected if they contain identified
electrons or muons or if any jet or $\tau$ candidate is consistent with arising from
detector noise or non-collision background~\cite{jetcalib}. Events are required to contain
one or more identified $\tau$ candidates, at least two jets,
one with $\pt>30\gev$ and another with $\pt>130\gev$, and missing
transverse momentum $\met>130\gev$.
The latter two requirements ensure that the trigger efficiency
is above 98\% in both data and simulation.

The two jets leading in \pt\ are required to be separated in azimuth from the
direction of the missing transverse momentum by more than $0.3$~rad. This requirement
reduces multijet events, which typically have instrumental missing
transverse momentum aligned with the leading jets. Multijet events are further suppressed
by requiring $\met/m_\mathrm{eff}>0.25$, where the effective mass, $m_\mathrm{eff}$, is
defined as the scalar sum of \met, the \pt\ of the two leading jets, and the
$\pt^\tau$ of the leading $\tau$ candidate.

Events are required to have a transverse mass, \mtrans, above $110\gev$.
The transverse mass is defined as
\begin{eqnarray}
\mtrans=\sqrt{m_\tau^2 + 2\pt^\tau\met\left(1-\cos\Delta\phi(\pt^\tau,\met)\right)}, \nonumber
\end{eqnarray}
where $\Delta\phi(\pt^\tau,\met)$ is the azimuthal angle between the $\tau$
and the direction of the missing transverse momentum.
This requirement suppresses backgrounds due to \Wjets
and top-quark production. The remaining SM backgrounds are
further suppressed by requiring $m_\mathrm{eff}>600\gev$.
This is the final selection defining the signal region for the analysis.
The \mtrans\ and $m_\mathrm{eff}$ requirements as well as the criteria used for
the suppression of multijet events are chosen to maximise
the signal significance computed with the Asimov approximation~\cite{Asimov}.

\section{Background estimation}

Background processes are divided into three classes which
are estimated separately: events with true $\tau$ leptons from
$t\to b\tau\nu$ decays (both top-quark-pair and
single top quark production) and $\Wboson(\to\tau\nu)+\mathrm{jets}$
events; events with misidentified (`fake`) $\tau$ candidates in top, \Wjets,
and \Zjets events; and events with fake $\tau$
candidates in multijet events. The two fake-$\tau$ classes are
treated separately to account for differences in $\tau$ misidentification 
probabilities due to different event topologies and jet composition.

Events with true $\tau$ leptons are
estimated in a control region defined by replacing
the requirement on the transverse mass in the final selection with the 
requirement $\mtrans<70\gev$.
For events with a correctly reconstructed $\tau$ lepton and
with \met\ entirely due to a single neutrino, \mtrans\ is
kinematically bounded from above by the \Wboson\ mass, within the detector resolution;
by requiring $\mtrans<70\gev$, more than 90\% of the events
in the resulting control region are expected to contain
true $\tau$ leptons from top-quark and $\Wboson$ decays.
The composition of the event sample in this control region is given
in Table~\ref{tab:CR_TrueTau}.
Within this control region, the background due to \Zboson\ decays
is estimated from simulation and the remaining small background
due to multijet events is estimated using a procedure similar
to that used to estimate the multijet background in the signal region,
described below.

Within the $\mtrans<70\gev$ control region,
top-quark and \Wjets yields are estimated individually with a 
maximum-likelihood fit
to the output distribution of a BDT built from four variables:
the number of $b$-quark jets, the total jet multiplicity,
the transverse momentum of the second-leading jet, and
the transverse thrust~$T$ of the event, defined as
$T = {\rm max}_{\hat{n}} \{ \sum_i{\hat{n} \cdot \vec{p}_{\mathrm{T},i}}/|\sum_i{\vec{p}_{\mathrm{T},i}}| \}$,
where $i$ runs over the missing transverse momentum and all jets, excluding the tau candidates, with transverse momentum vectors $\vec{p}_{\mathrm{T},i}$,
and the transverse thrust axis is given by the unit vector $\hat{n}$ for which the maximum
is attained. Top-quark events have more reconstructed $b$-quark jets, 
a higher jet multiplicity, higher jet momenta,
and tend to be more spherical than \Wjets events. 
Jets containing $b$ quarks are identified with about 60\% efficiency, evaluated with top-quark events,
using secondary vertex reconstruction and three-dimensional impact
parameters of tracks associated with the jet~\cite{btagging}.
The output distribution of this BDT is shown in Figure~\ref{fig:bdt}
along with the results of the fit. The results of the fit are scale
factors for \Wjets and top quark backgrounds which reflect
differences in cross sections and reconstruction efficiencies between
data and simulation. The measured scale factors are $1.22\pm 0.13$
for top events and $0.71\pm 0.03$ for \Wjets events. These scale
factors are applied to simulated event samples in the signal region
to derive the final expected true-$\tau$ yields from background processes.

\begin{table}[hbt]
\caption{Numbers of observed and expected events in the true-$\tau$-dominated
\Wboson/top control region, defined as $\mtrans<70\gev$. The numbers shown for \Wjets and
top are from Monte Carlo simulation and do not include the correction factors
derived from this control region. The correction factors obtained from
a fit to data are $1.22 \pm 0.13$ for top and $0.71 \pm 0.03$ for \Wjets. 
The true-$\tau$ purity is $97\%$ for top, $96\%$ for \Wjets and $87\%$ for \Zjets.}
\label{tab:CR_TrueTau}
\begin{center}
\begin{tabularx}{\columnwidth}{@{\extracolsep{\fill}}ccccc}\hline\hline
 Top              &\Wjets & \Zjets & Multijet             &Data  \\ \hline
 186.4 $\pm$ 8.4  & 919 $\pm$ 40           & 62.2 $\pm$ 6.7          & 1.8 $\pm$ 1.8 & 951  \\ \hline\hline
\end{tabularx}
\end{center}
\end{table}

\begin{figure}[hbt]
\centering
\includegraphics[width=0.49\textwidth]{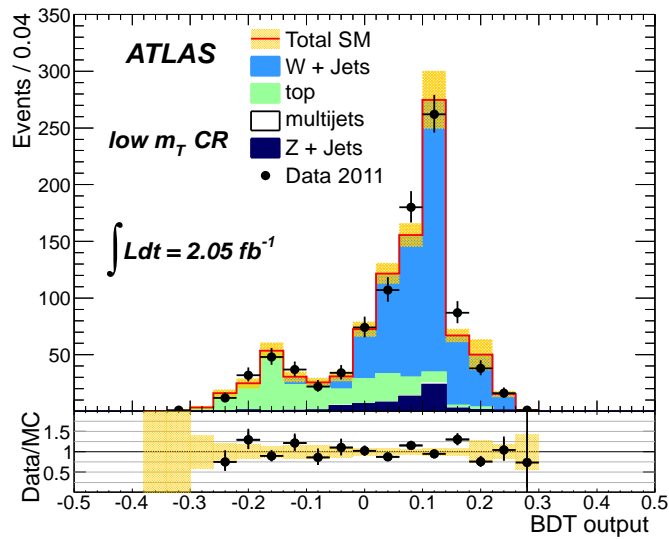}
\caption{Output distribution of the BDT used to discriminate \Wjets
from top-quark events in the low-\mtrans control region, defined as $\mtrans<70\gev$. Background
distributions are taken from simulation. The yield for \Wjets
and top backgrounds are taken from a maximum-likelihood fit to this
distribution. The solid (red) line with shaded (yellow) error band corresponds to the
total SM prediction, while the points are data.}
\label{fig:bdt}
\end{figure}

For the estimation of backgrounds due to fake $\tau$ candidates in top-quark,
\Wjets, and \Zjets events, a
second control sample is defined by selecting events that fulfill the event selection 
but with modified criteria on
\mtrans\ and $m_\mathrm{eff}$: $\mtrans > 70 \gev$ and either $\mtrans < 110 \gev$ 
or $m_\mathrm{eff} < 600 \gev$.
Since the \mtrans distribution falls off rapidly above the \Wboson\
mass for true-$\tau$ events, the intermediate \mtrans region selected
here is relatively enhanced in fake-$\tau$ events, and the overall
composition of this region is expected to be very similar to that of the
signal region.
Multijet events are expected to make up less than 3\% of this
sample and are estimated from simulation.
The composition of the fake-$\tau$-enhanced sample in this control region is shown in 
Table~\ref{tab:CR_FakeTau}.
Within this control region, true-$\tau$ backgrounds are subtracted
using estimates derived from the true-$\tau$-dominated control region.
The numbers of events remaining after the true-$\tau$ subtraction are used to
determine a scale factor, $0.50\pm 0.08$, which is
then applied to simulated samples of
fake-$\tau$ events in the signal region to obtain a final background
estimate. While this scale factor differs significantly from unity,
it is consistent with other ATLAS studies of the performance of
$\tau$ fake rates in simulation.

\begin{table}[hbt]
\caption{Numbers of observed and expected events in the fake-$\tau$-enhanced
control region. The numbers
of expected \Wjets and top-quark events have been corrected 
by the correction factors measured in the true-$\tau$-dominated region.
The fake-$\tau$ correction factor obtained from data is $0.50\pm 0.08$.}
\label{tab:CR_FakeTau}
\begin{center}
\begin{tabular}{l| r@{ $\pm$ }l r@{ $\pm$ }l| r@{ $\pm$ }l}\hline\hline
\ & \multicolumn{2}{c}{True $\tau$} & \multicolumn{2}{c}{Fake $\tau$} & \multicolumn{2}{c}{Total}\\\hline
Top                     & 53.3    & 7.5     & 37.8    & 5.8     & 91.1    & 9.4 \\
\Wjets & 80.5    & 6.9     & 33.3    & 4.1     & 113.8   & 8.0  \\
\Zjets & 5.1     & 1.6     & 41.5    & 10.8    & 46.6    & 10.9 \\
Multijet                & 0       & 0       & 2.9     & 1.0     & 2.9     & 1.0  \\\hline
Total                   & 139   & 10    & 116   & 13    & 254   & 17 \\\hline
Data                    & \multicolumn{4}{c}{}     & \multicolumn{2}{c}{197}\\\hline\hline
\end{tabular}
\end{center}
\end{table}

Backgrounds due to multijet events are estimated in a third control
region in which either $\met/m_\mathrm{eff}<0.25$ or one of the two
leading jets is aligned in
azimuth with the missing transverse momentum direction.
Within this sample, the probability
for jets (which contain very few true $\tau$ leptons) to satisfy
the $\tau$ selection criteria is estimated by applying the selection to
randomly chosen jet candidates. This probability is then applied
to a complementary sample
of multijet events, where the azimuthal separation and
$\met/m_\mathrm{eff}$, as well as all other event selection requirements, match those of the signal region,
but where the $\tau$ candidate is again randomly chosen from among the jet candidates. This provides 
an estimate of the multijet background yield in the
signal region. It is found that the multijet background 
makes up only a few percent of the total SM background 
in the signal region. 

Possible contamination from SUSY signals has been considered in
all three background-estimation control regions and is found
to have a negligible effect on the results presented below.

\section{Systematic uncertainties}

Dominant systematic uncertainties on the estimated background yields
are due to uncertainties in the jet energy scale (3--8\%)~\cite{jetcalib}, jet energy resolution (6--13\%)~\cite{jetcalib},
$\tau$ energy scale (2--10\%)~\cite{tauid},
statistical uncertainties in the data control regions (5--15\%),
and Monte Carlo uncertainties related to the extrapolation from
the control regions to the signal region (10--20\%).
This last term includes statistical uncertainties in the simulation,
variations in the in the assumed \Wjets/top/\Zjets mixture in the fake-$\tau$ control region, and
Monte Carlo generator uncertainties (estimated by varying the shower matching,
factorization and renormalization scales, $\alpha_s$, and the amount of initial-state and final-state radiation)~\cite{susy-b-jet}.
Additional uncertainties on \Wjets and top-quark
backgrounds are estimated by varying the assumed $b$-quark
identification efficiency within measured uncertainties (4--11\%)~\cite{btagging}.
Uncertainties on the multijet background yield are estimated
by studying correlations between $m_\mathrm{eff}$ and the
azimuthal separation between the leading two jets and the missing
transverse momentum. Additional systematic uncertainties, including
those on the pile-up description in the simulation, are considered and
found to be negligible.

In addition to the sources described above, systematic uncertainties on the SUSY signal cross section are estimated by
varying the factorization and renormalization scales in \texttt{PROSPINO}
up and down by a factor of two, by considering variations in $\alpha_S$, and by varying the proton PDFs within their uncertainties.
These theoretical uncertainties total
typically 8--12\% across the relevant region of parameter space.
Uncertainties are calculated separately for individual SUSY
production processes.

\section{Results}

Figure~\ref{fig:dataMC}
shows the distributions of \met, $\pt^\tau$, and $m_\mathrm{eff}$
for data with all selection requirements applied except for that on
$m_\mathrm{eff}$, along with the corresponding estimated backgrounds.
The numbers of expected SM background events and the observed
number of events after the $m_\mathrm{eff}$ requirement are shown in Table~\ref{tab:yields}. The data
agree with the background expectation. 

Based on these results, limits are placed on
contributions beyond the SM to the signal region. With 11 events observed and
$13.2\pm 4.2$ expected, an upper limit of $8.5$ on the number of events
observed due to non-SM sources is derived at
95\% confidence level (C.L.). This limit corresponds to
an upper limit on the visible cross section of $4.0\ \mathrm{fb}$,
where the visible cross section is
defined as the product of production cross section,
branching fraction to at least one $\tau$ lepton, acceptance, and
efficiency using the event selection defined in Section~\ref{sec:evsel}.
For the two benchmark points $\Lambda = 30$, $\tan\beta = 20$ and $\Lambda = 40$, $\tan \beta=30$
the product of branching ratio to $\tau$-leptons, the acceptance and the efficiency 
for this selection amounts to $1.47\%$ and $1.69\%$, respectively.
Figure~\ref{fig:limit} shows an interpretation of the result as
a 95\% C.L. exclusion limit in the $\mmess=250\tev$,
$N_5=3$, $\mu>0$, $C_\mathrm{grav}=1$ slice of the GMSB model.
Figure~\ref{fig:limit} also shows the variation of the expected limit in
response to $\pm 1\sigma$ fluctuations in the expected SM background and
the SUSY cross sections.
The excluded regions are calculated using a profile likelihood method
with systematic uncertainties modelled as varying Gaussian-distributed
nuisance parameters~\cite{limits,cls}.
The resulting limit is compared with previous exclusion limits from
searches for $\tilde{\tau}$ and $\tilde{e}$ production and GMSB topologies at LEP.
The region of small $\Lambda$ and large $\tan\beta$ is theoretically
excluded since it leads to tachyonic states.
In this model, the production of supersymmetric particles can be
excluded at 95\% C.L. up to $\Lambda=30\tev$, independent of $\tan\beta$,
and up to $\Lambda=43\tev$ for large values of $\tan\beta$.

\begin{table*}[hbt]
\caption{Expected SM background event yields and number of events
observed in data after the final requirement on $m_\mathrm{eff}$.
All systematic uncertainties are included here,
and the uncertainty on $\Sigma_\mathrm{SM}$, the sum of all SM
backgrounds, takes correlations between the individual background
uncertainties into account. The true-$\tau$ purities 
are $53\%$ and $64\%$ for the top and \Wjets backgrounds, respectively,
and are negligible for the \Zjets and multijet backgrounds.  
For comparison, the estimated event yield for a GMSB
signal with $\Lambda=40\tev$, $\tan\beta=30$ is an additional $9.1\pm 1.7$ events.}
\label{tab:yields}
\begin{center}

\begin{tabular}{cccccc}\hline\hline
Top          & \Wjets       & \Zjets       & Multijet     & $\Sigma_\mathrm{SM}$ & Data      \\\hline
$5.6\pm 1.4$ & $4.7\pm 1.5$ & $2.4\pm 0.7$ & $0.5\pm 0.6$ & $13.2\pm 4.2$        & 11        \\\hline\hline
\end{tabular}

\end{center}
\end{table*}

\begin{figure}[hbt]
  \centering
  \subfigure{
    \includegraphics[width=0.45\textwidth]{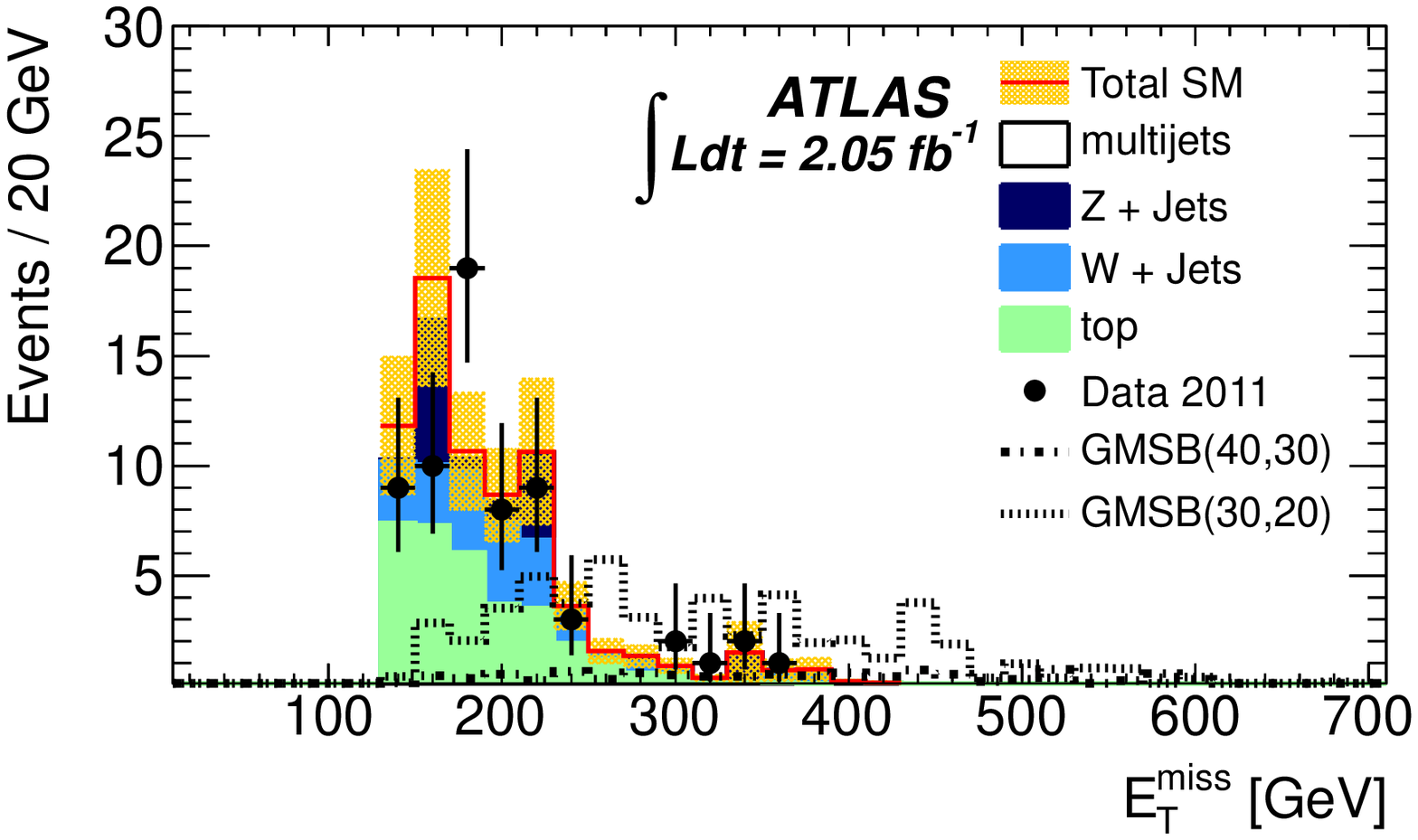}
  }
  \subfigure{
    \includegraphics[width=0.45\textwidth]{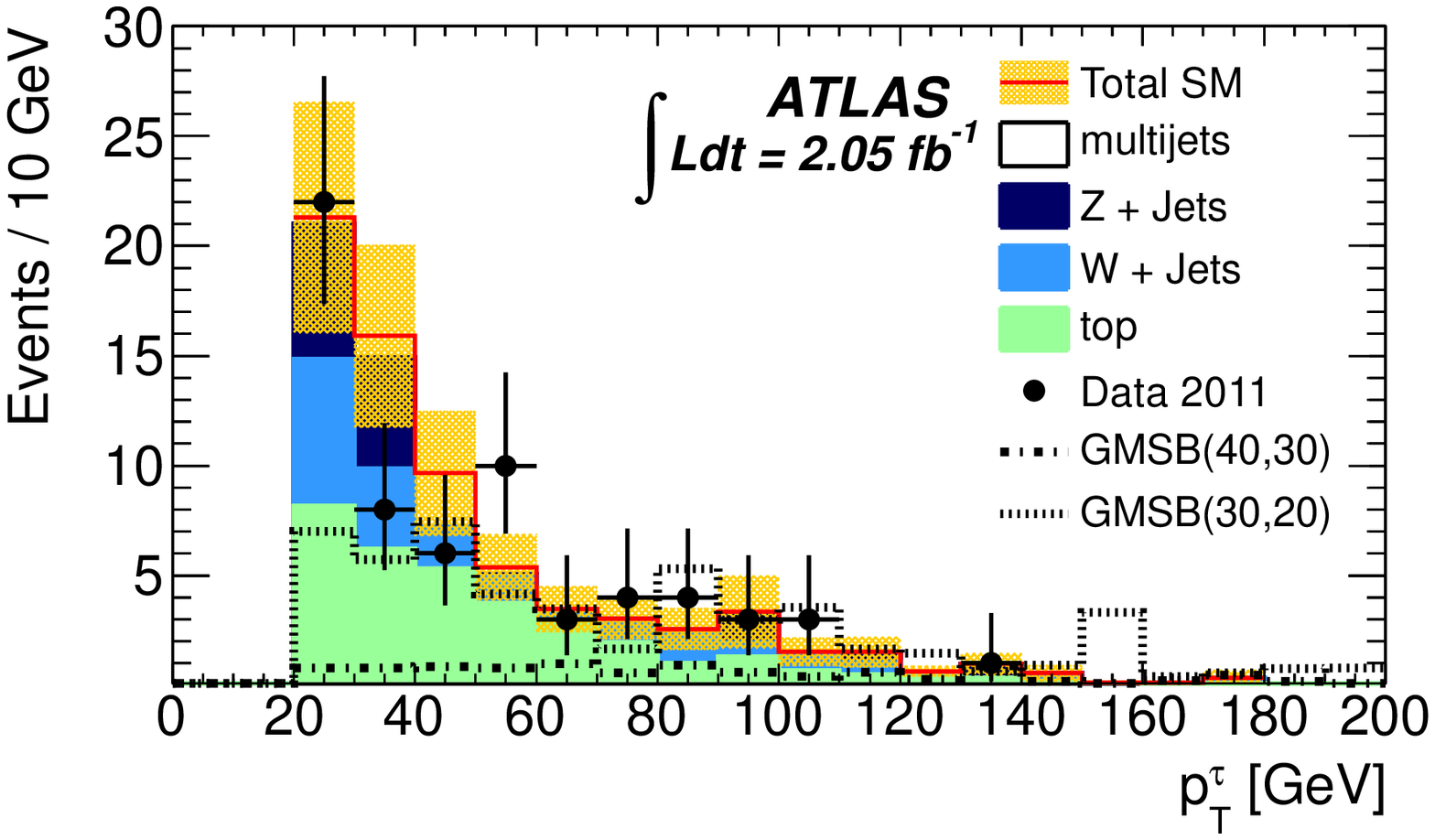}
  }
  \subfigure{
    \includegraphics[width=0.45\textwidth]{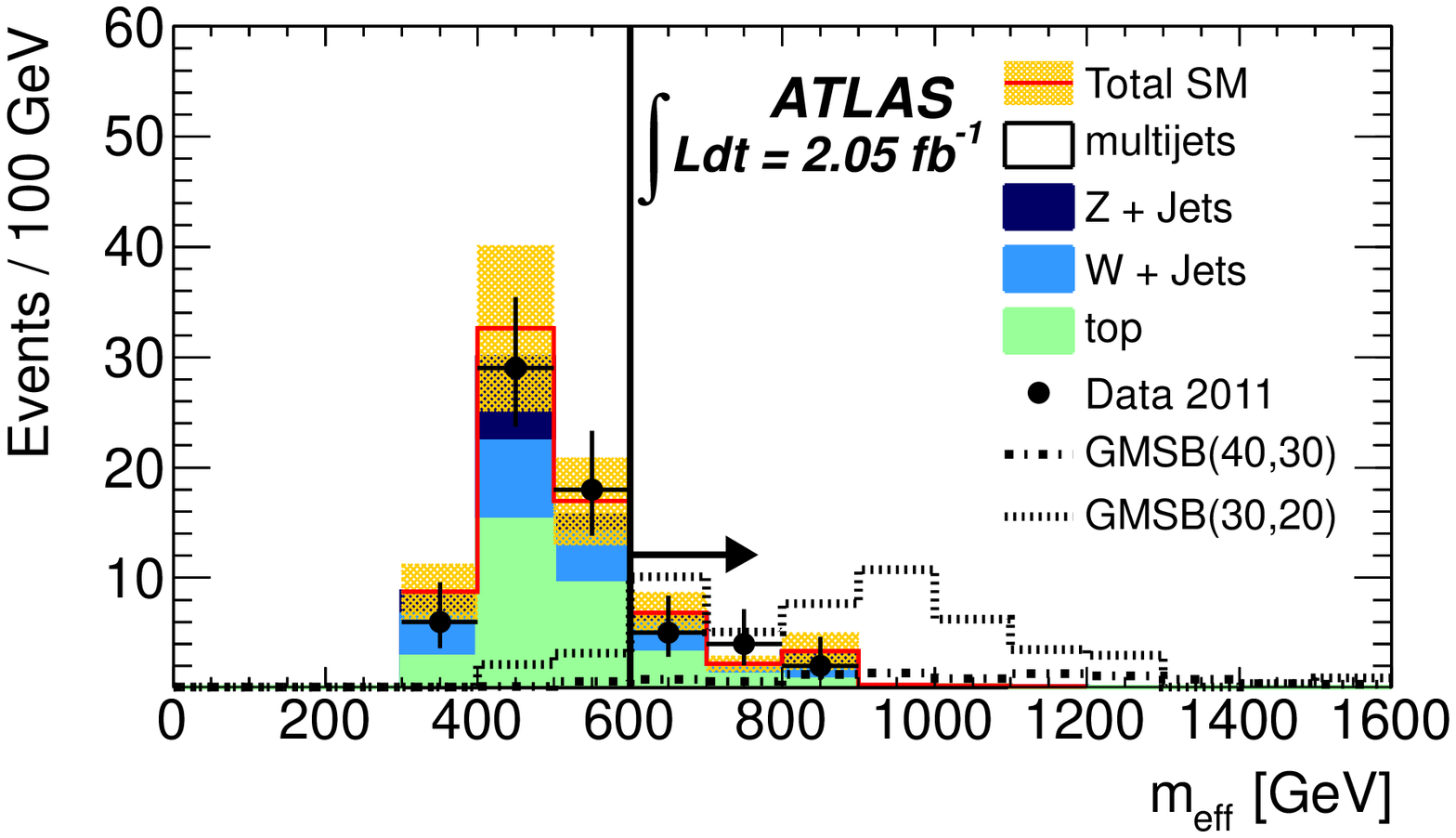}
  }
\caption{Distributions of \met, $\pt^\tau$, and $m_\mathrm{eff}$ for
data with all selection requirements except for that on $m_\mathrm{eff}$, 
along with the corresponding estimated backgrounds.
Backgrounds are taken from simulation and normalised with control regions in data.
The solid (red) line with shaded (yellow) error band corresponds to the
total SM prediction, while the points are data.
The error bands indicate the size of the total (statistical and systematic) uncertainty.
The notation GMSB(40,30) stands for the GMSB model with $\Lambda=40\tev$ and $\tan\beta=30$ and analogously for GMSB(30,20).}

\label{fig:dataMC}
\end{figure}

\begin{figure}[hbt]
\centering
\includegraphics[width=0.48\textwidth]{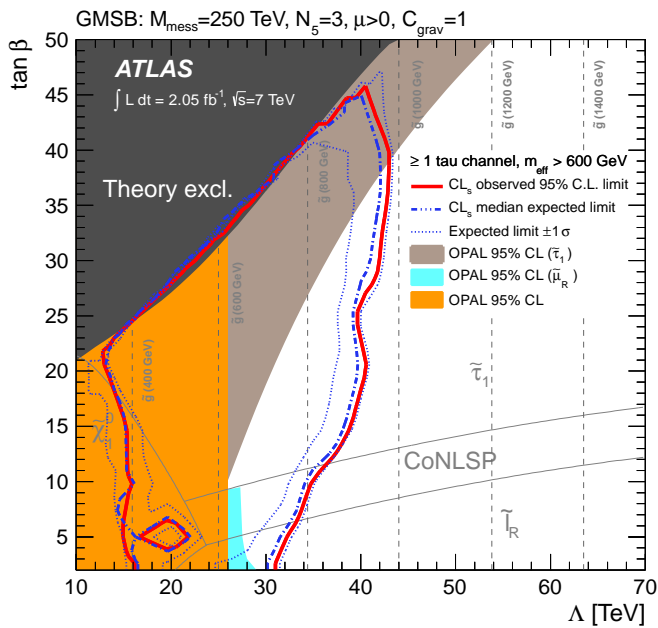}
\caption{Expected and observed 95\% C.L. exclusion limits in the $\mmess=250\tev$,
$N_5=3$, $\mu>0$, $C_\mathrm{grav}=1$ slice of GMSB, together with
the most stringent previous limits from OPAL~\cite{opal}. The identity of the NLSP is indicated, with
CoNLSP the region where the $\tilde{\tau}$ and $\tilde{\ell}$ are nearly degenerate.}
\label{fig:limit}
\end{figure}

\section{Conclusions}

In conclusion, this Letter presents a search for supersymmetry in final states
containing jets, missing transverse momentum, and at least one
$\tau$ lepton with the ATLAS experiment in $\sqrt s=7\tev$ proton-proton collisions 
at the LHC.
This is the first search in these final states at the LHC that includes events with 
one $\tau$ lepton.
No excess of events is seen beyond the expected Standard Model backgrounds in $2.05\ \ifb$ of data.
Limits are placed on the visible cross section and in the context of GMSB models. 
The limits obtained extend the results from previous experiments.

\section{Acknowledgments}

We thank CERN for the very successful operation of the LHC, as well as the
support staff from our institutions without whom ATLAS could not be
operated efficiently.

We acknowledge the support of ANPCyT, Argentina; YerPhI, Armenia; ARC,
Australia; BMWF, Austria; ANAS, Azerbaijan; SSTC, Belarus; CNPq and FAPESP,
Brazil; NSERC, NRC and CFI, Canada; CERN; CONICYT, Chile; CAS, MOST and NSFC,
China; COLCIENCIAS, Colombia; MSMT CR, MPO CR and VSC CR, Czech Republic;
DNRF, DNSRC and Lundbeck Foundation, Denmark; EPLANET and ERC, European Union;
IN2P3-CNRS, CEA-DSM/IRFU, France; GNAS, Georgia; BMBF, DFG, HGF, MPG and AvH
Foundation, Germany; GSRT, Gerece; ISF, MINERVA, GIF, DIP and Benoziyo Center,
Israel; INFN, Italy; MEXT and JSPS, Japan; CNRST, Morocco; FOM and NWO,
Netherlands; RCN, Norway; MNiSW, Poland; GRICES and FCT, Portugal; MERYS
(MECTS), Romania; MES of Russia and ROSATOM, Russian Federation; JINR; MSTD,
Serbia; MSSR, Slovakia; ARRS and MVZT, Slovenia; DST/NRF, South Africa;
MICINN, Spain; SRC and Wallenberg Foundation, Sweden; SER, SNSF and Cantons of
Bern and Geneva, Switzerland; NSC, Taiwan; TAEK, Turkey; STFC, the Royal
Society and Leverhulme Trust, United Kingdom; DOE and NSF, United States of
America.

The crucial computing support from all WLCG partners is acknowledged
gratefully, in particular from CERN and the ATLAS Tier-1 facilities at
TRIUMF (Canada), NDGF (Denmark, Norway, Sweden), CC-IN2P3 (France),
KIT/GridKA (Germany), INFN-CNAF (Italy), NL-T1 (Ne\-ther\-lands), PIC (Spain),
ASGC (Taiwan), RAL (UK) and BNL (USA) and in the Tier-2 facilities
worldwide.

\bibliographystyle{elsarticle-num}
\bibliography{SusyTau}

\onecolumn
\begin{flushleft}
{\Large The ATLAS Collaboration}

\bigskip

G.~Aad$^{\rm 48}$,
B.~Abbott$^{\rm 110}$,
J.~Abdallah$^{\rm 11}$,
S.~Abdel~Khalek$^{\rm 114}$,
A.A.~Abdelalim$^{\rm 49}$,
A.~Abdesselam$^{\rm 117}$,
O.~Abdinov$^{\rm 10}$,
B.~Abi$^{\rm 111}$,
M.~Abolins$^{\rm 87}$,
O.S.~AbouZeid$^{\rm 157}$
% ...
\bigskip

% ...
$^{10}$ Institute of Physics, Azerbaijan Academy of Sciences, Baku, Azerbaijan\\
$^{11}$ Institut de F\'isica d'Altes Energies and Departament de F\'isica de la Universitat Aut\`onoma  de Barcelona and ICREA, Barcelona, Spain\\
% ...
$^{48}$ Fakult\"{a}t f\"{u}r Mathematik und Physik, Albert-Ludwigs-Universit\"{a}t, Freiburg i.Br., Germany\\
$^{49}$ Section de Physique, Universit\'e de Gen\`eve, Geneva, Switzerland\\
% ...
$^{87}$ Department of Physics and Astronomy, Michigan State University, East Lansing MI, United States of America\\
% ...
$^{110}$ Homer L. Dodge Department of Physics and Astronomy, University of Oklahoma, Norman OK, United States of America\\
$^{111}$ Department of Physics, Oklahoma State University, Stillwater OK, United States of America\\
% ...
$^{114}$ LAL, Univ. Paris-Sud and CNRS/IN2P3, Orsay, France\\
% ...
$^{117}$ Department of Physics, Oxford University, Oxford, United Kingdom\\
% ...
$^{157}$ Department of Physics, University of Toronto, Toronto ON, Canada\\
% ...
\end{flushleft}



\end{document}